\documentclass[aps,prc,twocolumn,groupedaddress,showpacs,epsf]{revtex4}
\usepackage{epsfig}
\usepackage{graphicx}	
\newcommand{\cA}{{\cal A}}
\newcommand{\be}{\begin{equation}}
\newcommand{\ee}{\end{equation}}
\newcommand{\ba}{\begin{array}}
\newcommand{\ea}{\end{array}}

\begin{document}

\title{
Temperature dependence of nuclear matter generalized isovector symmetry energy
with Skyrme-type interactions}
\author{ F\'abio L. Braghin}
\affiliation{
 International Center for Condensed Matter Physics, 
Universidade de Brasilia, 
 Caixa Postal 04513,  70904-970, Brasilia, DF, Brazil\\
and
\\
Instituto de Fisica da Universidade de S\~ao Paulo, 
CP 66318, 05314-970, S\~ao Paulo, Brazil.}

\begin{abstract} 
The temperature dependence of the 
nuclear matter isovector symmetry energy coefficient ($\cA_{0,1}$) is investigated in the framework of the generalized nuclear polarizability with Skyrme interactions, as worked out in Refs.~\cite{npa,prc}. 
The  variation of $\cA_{0,1}(T)$ 
is very small (of the order of $1$ MeV) for temperatures (T)
in the range of $0$ and $18$ MeV. 
Different behaviors with 
temperature are found strongly depending on the Skyrme
 parameterization, in particular at  densities lower than the saturation density $\rho_0$.
\end{abstract}


\maketitle

\section{Introduction}

The neutron-proton nuclear symmetry energy is currently under intense theoretical and experimental investigations not only due to its relevance to different aspects of the nuclear structure and dynamics but also to description of dense stars structure and of the Supernovae mechanism \cite{shetty-0704,MSU-GSI,BKB,NN2006,GAITANOS,eos,SCALING-FRAG,npa,ASTRO}. 
Actually, the possibility of extracting better 
experimental values for it at different densities and temperatures
is under constant improvement. 
In this respect, (multi)fragmentation processes in heavy ion collisions provide
experimental knowledge about the liquid-gas phase transition and therefore 
about the nuclear forces at low nuclear densities and excitation energies.
For example, the isoscaling  found in yields of multi-fragmentation experiments \cite{SCALING-FRAG}
 depends directly on the  isovector 
symmetry energy (coefficient) which was found to decreases considerably with the 
excitation energy according to experimental data \cite{kowalski-etal}. 
In fact, in experimental conditions it is very difficult to extract unambiguous behavior 
 with each of the thermodynamic observables involved ($T,\rho$) 
\cite{kowalski-etal} and in 
earlier descriptions of experimental results some groups
have considered a seemingly too strong variation of the symmetry 
energy with temperature \cite{botvina-etal}. 
In different (more recent) analysis it has been found that 
the strong decrease of the symmetry energy in experimental 
conditions should also be due to to 
the excitation energy dependence of surface effects 
\cite{souza-etal} and
to the expansion of the system which implies the lowering the total density \cite{samaddar-etal,prc}. 
Besides that, recently De and Samaddar \cite{de-samaddar}  have argued  that 
the symmetry free energy 
\cite{NPA776} is the parameter that 
appears in the scaling of multi-fragmentation.
 However their arguments stand for the specific analysis of 
multi-fragmentation processes and we intend to discuss rather the behavior the symmetry energy coefficient behavior.
Concerning the specific dependence on the temperature (up to $T \simeq 20$MeV)
there are several theoretical indications of very weak decrease
 \cite{moustakidis07,dean95-etal,dean02-etal,xu06-etal,NPA776} 
although in some works, for finite nuclei and nuclear matter,
 a  small (or very small) increase was found
\cite{donati94-etal,dean95-etal,npa}.
In some of these works no unique tendency was found 
mainly for different
 finite nuclei, for example in Ref.\cite{dean95-etal}.
In Ref.\cite{npa}  few preliminary
numerical results
showed a very small variation of the nuclear matter symmetry energy coefficient 
with temperature
using Skyrme forces. In the present work we perform a quite extensive 
investigation
of this subject with the generalized polarizability as proposed 
in Refs.\cite{npa,prc}.

In usual mass formulae the binding energy per nucleon 
depends on the n-p asymmetry with the following form:
\be \ba{ll} \label{1}
\displaystyle{ \frac{E}{A} = \frac{H_0(A,Z)}{A} + a_{\tau} \frac{(N-Z)^2}{A^2} + 
. . .
}
\ea \ee
where $N,Z,A$ are the neutron, proton and mass numbers, the isovector symmetry 
energy coefficient (s.e.c.) is $a_{\tau}$.
This coefficient is a measure of the energy needed to increase n-p asymmetry.
Different powers of the n-p asymmetry $(N-Z)^j/A$  ($j \neq 2$) are quite smaller,
although the n-p asymmetry modifies many other terms \cite{MOLLER-NIX}.
The value of the n-p s.e.c. 
in fits of the mass formulae for stable nuclei is quite well known
\cite{shetty-0704}.
For an infinite medium, roughly speaking, nucleon numbers ($N,Z$) might be replaced by the corresponding densities $\rho_n,\rho_p$ for many purposes depending on  
the volume occupied by each nucleon species,
 see for example Ref.~\cite{NEUTRONdensity}.

A particular interesting way of obtaining the symmetry energy coefficient,
eventually in different conditions, 
was found by the nuclear isovector polarizability $\cA_{0,1}$ \cite{BVA,npa2001}.
Generalized nuclear polarizabilities and their dependencies
on several parameters simultaneously 
(such as total density $\rho$, exchanged momentum and energy $q, \omega$, 
neutron-proton asymmetry $b = \frac{\rho_{n}}{\rho_{p}} -1$) were investigated 
quite extensively with Skyrme-type forces 
\cite{npa2001,prc,flb-aip,rtfnb-02-07}. 
Deviations from the quadratic form of expression (\ref{1}) might appear depending on 
the particular density fluctuations considered  for a given set of 
nucleon densities for these calculations with Skyrme density functionals.
 Recent investigations have revealed further relations of 
these (non relativistic) 
effective interactions with 
relativistic models \cite{qmc-etc,cp-etal}.

This work exhibits the temperature dependence of the isovector 
generalized s.e.c., as proposed in Refs.~\cite{npa,prc} using 
Skyrme type forces. 
For that, the generalized isovector screening function with different parameterizations of
 Skyrme effective interactions are shown at different densities and n-p asymmetries.

\section{ Generalized  polarizabilities }

For an asymmetric medium in n-p densities, 
the generalized screening function using Skyrme forces, for  zero energy and momentum exchange
($\omega = q = 0$), can be written in a compact notation, as \cite{npa,prc}:
\begin{eqnarray} \label{20} 
\cA_{s,t} = 
\frac{\rho }{ 2 N }
\left\{ 1 + 2\overline{V_0^{(s,t)}} N_c 
+6 V_1^{(s,t)}M_p^* ( {\rho}_c  + {\rho}_d ) + \right. 
\nonumber
\\
\left. +
12M^*_p V_1^{(s,t)} \overline{V_0^{(s,t)}} \left( N_c {\rho}_d 
 - {\rho}_c  {N}_d  \right) + \right.  \\
\nonumber
\left. + (V_1^{(s,t)})^2 \left( 36(M^*_p)^2 {\rho}_c 
{\rho}_d  - 16M_p^* M_c  N_d \right)  \right\} .
\nonumber
\end{eqnarray}
Where $\overline{V_0}^{(s,t)} $ and $V_1^{(s,t)}$ are functions 
of the Skyrme forces parameters in a given channel 
of the effective nuclear interaction
of (spin, isospin) denoted
by superscripts ($s,t$).
Therefore they carry the main contributions of the effective NN interaction,
 distinguishing each of the channel of the particle-hole interaction. 
In particular, the function $V_1^{(0,1)} = (t_2(1+2x_2)-t_1(1+2x_1))/16$ is
 a combination of (momentum dependent) 
Skyrme parameters that contributes to the usual nucleon 
effective masses in the framework of Skyrme  calculation, $m^*_{n,p}$.
The effective masses of neutron and protons are
functions of the  total density and of the
 neutron and proton densities, 
see for example in \cite{BVA}. 
On the other hand, the function $V_0^{(0,1)}$ depends mainly on $t_0$ and $t_3$.
While $V_0^{(0,1)}$ of each of the Skyrme parameterization used in this work do not 
have meaningful different values, the function $V_1^{(0,1)}$ has considerably different 
values because it might be zero, i.e. for some forces $V_1^{(0,1)} = 0$. 
Although relevant these differences will be shown to not be enough to provide too much
 different behaviors for $\cA_{s,t} (T)$.
The mixed functions $\rho_v $, $M_v $ and $N_v $ reduce to densities of the Skyrme-Hartree-Fock approach
 at zero temperature, being respectively the nucleon and 
kinetic energy densities and the densities of states.
They are the zero frequency and zero momentum  generalized 
Lindhard functions calculated in Ref.~\cite{npa}. 
The total densities, $\rho, N$, are written without any index.
These functions ($\rho_v, M_v, N_v$) are given respectively by:
\be \ba{ll} \label{densities}
\rho_v = v \rho_n + (1-v) \rho_p, \\
 M_v  = v M_n  + (1-v) M_p, \\
 N_v  = v N_n  + (1-v) N_p.
\ea \ee
In these expressions $v$ stands for two different n-p asymmetry coefficients ($c,d$), being that we made use of the 
 following asymmetry coefficients: 
$$
a = \frac{m^*_{p}}{m^*_{n}} - 1 , \;\;\;\;
b = \frac{\rho_{n}}{\rho_{p}}  -1, \;\;\;\; 
c = \frac{1+b }{2+b }, \;\;\;\;
d = \frac{1 }{1+ (1+b)^{\frac{2}{3} } }.
$$
Therefore, by fixing the parameter of density asymmetry, $b$ 
(for a given total nuclear density),
the neutron and proton densities
 are found as well as 
the other asymmetry parameters ($a,c,d$) for a given Skyrme force
\cite{npa,prc}.

There is a further mass parameter  in expression (\ref{20}), $M^*_p$ which is in fact a kind of reduced mass. It is given by
$$
M^*_p \equiv \frac{m_p^*}{ (1 + \frac{a}{2}) } = 
\frac{2 m_p^* m_n^*}{ m_p^* + m_n^*}.
$$
Although the calculation has been carried out in such a way to provide final expressions for each of the channels of the particle-hole interaction (isovector, spin, spin-isovector and scalar) only the n-p one
will be investigated in this paper.

\subsection{Varying Temperature}

The densities $N_{\alpha}, \rho_{\alpha}$ and $M_{\alpha}$ (for $\alpha = n,p$ neutrons and protons)
are the basic input for the temperature dependence of the polarizabilities.
At finite temperature these functions 
 are given respectively by  integrals written as:
\begin{eqnarray} \label{densities-int}
(N_{\alpha}, \; 3 \rho_{\alpha}, \; 4 M_{\alpha}) 
= 
- \frac{1}{\pi^2}
\int d f_{\alpha}(k) (k . m^*_{\alpha}; \; k^3; \; k^5).
\end{eqnarray}
In these expressions, $d  f_{\alpha}(k)$ is the measure of integration 
in terms of the usual free fermion occupation numbers $f_{\alpha}(k)$  for neutrons  and protons ($\alpha=n,p$).
At $T=0$ for the usual  Fermi occupation number we have: 
\begin{eqnarray} \label{dfk}
d f_{\alpha}(k) = - \delta( k - k_F^{(\alpha)} ) dk,
\end{eqnarray}
Where $k_F^{(\alpha)}$ is the Fermi momentum for each of the nucleon species.
In this case the integration is trivial. Therefore all the temperature dependence 
of the density-like quantities $\rho,N,M$ is encapsulated in the integrals above
(\ref{densities-int}) which can, at most, yield smoother results for the integral.
Furthermore $N_v(T), M_v(T)$ are the only parameters that vary with temperature since $\rho_{\alpha}$ are kept constant.
The zero temperature limit 
was considered previously \cite{npa,prc} and it shows more explicitly, as mentioned above, the effect of each of the Skyrme force parameters through the functions $V_0^{(s,t)}$ 
and $V_1^{(s,t)}$ as well as the effective masses.
This issue is extremely relevant for the resulting $\cA_{0,1}(T)$.

From the general expression (\ref{20})
an useful (simplified) limit is recovered in which the behavior with temperature
 can be  understood in detail.
For instance, consider the limit in which the function $V_1^{(0,1)}$ appears only 
in the leading order of the symmetric n-p function.
This is achieved with $ b = a= 0$ and $d = c = 1/2$,  yielding
$m^* = m_n^* = m_p^* =  M_p^*$
 and $\rho_p = \rho_n = \rho/2$.
We obtain an expression of the following form, for $(s,t) = (0,1)$, in the n-p symmetric limit:
\begin{eqnarray} 
\label{A-tau}
\cA_{0,1} \to a_{\tau} = 
\frac{\rho}{2} \left(
\frac{ 1}{ N} + 2 V_0^{(0,1)} + 6 m^* V_1^{(0,1)} \frac{\rho}{N}
\right) + h.o.
\end{eqnarray}
Where
$h.o.$ stands for the higher order terms in $V_1^{(0,1)}$.
This expression reproduces exactly
$a_{\tau}$, which is the usual  symmetry energy coefficient
in the Skyrme-Hartree-Fock approximation \cite{BVA}. 
The first and third terms of $a_{\tau}$ depend on $N$ and therefore they
 show a very small variation with temperature only due to $N(T)$ 
according to expression (\ref{densities-int}).
 The  behavior of the function  $N_{\alpha}(T)$
is monotonic and it decreases with T. 
This is the main feature for understanding the numerical
results of expression (\ref{20}).

The higher order terms, in the limit of n-p symmetric matter, are given by:
\begin{eqnarray}
h.o. \to (V_1^{(0,1})^2 \left( 9 (m^*)^2 \rho^2 - 8 m^* M N \right)
\frac{\rho}{2 N}.
\end{eqnarray}
In these terms, and mainly for
 n-p asymmetric matter (in the complete expression (\ref{20})), 
the imbalance between the T-dependence of $M_{v}(T)$ and $N_v(T)$ determines 
whether the increasing behavior of $N(T)$ with temperature is the leading one or not.
Basically this is seen from the (overall) denominator of expression (\ref{20})
by reminding that $N(T)$ is a decreasing function of the temperature in the Skyrme-Hartree-Fock level, and on the other hand $M_v(T)$ is an increasing function of the temperature.
The different resulting behaviors might appear due to
 the relative values of their coefficients, i.e. $V_0^{(0,1)}$ and $V_1^{(0,1)}$.
This is noticed in the results exhibited in the next section.
Although it might be expected that for asymmetric n-p matter the variation of the polarizability is larger  since there are more terms at work, this 
will be shown to be not really sizeable for the forces considered in this work.
 The complete expression is quite complicated such that it might 
not exhibit a simple and unique behavior in more general situations.

It is worth to mention some recent results claiming that the symmetry free energy is the quantity that really rules multi-fragmentation 
\cite{NPA776,de-samaddar,samaddar-etal-08}. 
Basically this corresponds to considering the entropic contribution 
 which amounts basically to extra additive terms.
This can be qualitatively seen as follows.
The calculation of the polarizability in such case should depart from a free energy in the presence of a n-p asymmetry and of an infinitesimal 
external source ($\epsilon$) that induces fluctuations 
of the densities of neutrons from protons, i.e:
${\cal F } (\rho + \delta \rho_{np}) = {\cal E } (\rho + \delta \rho_{np})  - 
T S (\rho + \delta \rho_{np}) + \epsilon \delta \rho_{np},$
where $\delta \rho_{np} = \delta \rho_n - \delta \rho_p$.
The entropy can be expanded in terms of $\rho_n - \rho_p$
to make explicite its contribution to the symmetry (free) energy:
$
{\cal F } (T)  =  {\cal E}_0 (T) + a_{\tau} (\delta \rho_{np})^2  + 
S^{(1)} (T) \delta \rho_{np} - S^{(2)} (T) (\delta \rho_{np})^2 + \epsilon \delta \rho_{np}  + ..
$
Where $S^{(i)} (T)$ are the leading contributions of the entropy for the symmetry free energy.
As it was shown in Ref.~\cite{prc}, the linear term in $\delta \rho_{np}$ might be incorporated into the usual calculation (considering only the quadratic terms) and it is not considered explicitly 
below. 
The polarizability is then given by:
\begin{eqnarray} \label{entropic}
\Pi \equiv \frac{\delta \rho_{np}}{\epsilon}
 = - \frac{\rho}{2 (a_{\tau} - S^{(2)} ) }.
\end{eqnarray}
The
final symmetry free energy coefficient can be written 
as $a^{f}_{\tau} (T) = a_{\tau} (T) - S^{(2)} (T)$.
Therefore we can expect that the entropic contribution 
would appear mainly as additive terms for the screening function. 
This is seen in the results of 
Refs.~\cite{NPA776,samaddar-etal-08,de-samaddar}.
 However a microscopic investigation of this quantity, with its eventual relevance for the multi-fragmentation processes, is outside the scope of the present work.

\subsection{Results}

The temperature dependence of the isovector polarizability 
$\cA_{0,1} (T)$ is shown in figures 1 to 4 for the following Skyrme 
forces: SGII from Ref.~\cite{sgii}, SLyb from Ref.~\cite{sly}
(which is sometimes referred to as SLy4 in the literature) 
and two parameterizations SkCS4 and SkSC6 from Ref.\cite{sksc}.
These last two Skyrme parameterizations have a slightly more intricate 
density dependence although the resulting functions $V_1^{(0,1)}$ are zero.
The zero function $V_1^{(0,1)}=0$ 
(that carries the main part of the momentum dependent
Skyrme forces) brings a lot of simplification in the dependence on the 
temperature as discussed in the last section and it is noticed in the 
figures below. 
In these cases the behavior of $\cA_{s,t} (T)$ is always
monotonic depending on $N(T)$, and slightly less on $V_0^{(0,1)}$.
Different total densities and  n-p density asymmetries are also considered.

In Figure 1 the function $\cA_{0,1} (T)$ is shown at the saturation
density, $\rho = \rho_0$, and 
zero n-p density asymmetry ($b=0$) with the following Skyrme parameterizations: 
SkSC4  (circles) \cite{sksc}, SkSC6 ($\times$) \cite{sksc}, SGII (squares) \cite{sgii}, SLyb  (diamonds) \cite{sly}.
The variation with temperature (up to $T \simeq 18$ MeV) is quite small, 
reaching $\Delta \cA_{0,1} \simeq 0.5-1.0$ MeV depending on the interaction, and 
even nearly zero for the SGII force. 
The slope is always positive, although smaller at high temperatures.
As noticed after expression (\ref{A-tau}) the function $N(T)$ decreases with temperature within the Skyrme-Hartree-Fock approach.
The force SLyb is the one with larger variation in 
$\cA_{0,1}(T)$.
This trend of small variation was found before \cite{npa}, although it disagrees with the small decrease of the s.e.c. with $T$ found 
in different works for lower densities
\cite{moustakidis07,dean95-etal,donati94-etal,dean02-etal,xu06-etal,NPA776}.
However it is also worth to point out that in these references the s.e.c. was investigated in the regime of very low total density.
The  variation found for the range of $T=0$ up to 
$T\simeq 15$MeV is not large in all these works, 
and it can be of the order of 1 up to 3 MeV (though negative), a little bit larger than the present results.

Some further remarks to understand the behavior with temperature  are in order.
The chemical
 potential fixes the nucleonic density which is kept constant for 
all temperatures.
In the calculations with Skyrme interactions, 
the kinetic part of $a_{\tau}$ and the  terms with $N(T), M(T)$ in expression (\ref{20})) 
  are temperature-dependent. 
In particular whereas the functions $N_{\alpha}$ slightly 
decrease with the temperature, 
the densities $M_{\alpha}$ increase slightly. 
Having this in mind and analyzing expression (\ref{20}) we can 
expect that n-p asymmetry ($b, d, c \neq 0$) favors different behaviors 
of  $\cA_{0,1} (T)$. 
The relative variation of the potential energy part of the symmetry energy at the Hartree Fock level is nearly zero since it depends mostly on $\rho$ at not very low densities.

The same Skyrme parameterizations (and symbols) are used in Figure 2. 
In this figure, $\cA_{0,1} (T)$ is exhibited for $\rho = 0.75 \rho_0$ with $b=0.25$ 
(full symbols and $\times$) and $b=0$ (empty symbols and $+$).
For these cases (of lower nuclear matter density), the behavior of the polarizability with temperature is non-monotonic in the case of force SGII (squares).
For Skyrme forces SkSC4, SkSC6 and SLy, the polarizability $\cA_{0,1} (T)$ is nearly constant at very low temperatures and it starts increasing smoothly around $T \simeq 5$MeV until $T\simeq 10-15$MeV.
For the force SGII, $\cA_{0,1}$ decreases for low temperatures and smoothly increases for temperatures higher than nearly $3 -5$MeV.
By comparing the relative variation of the polarizability for 
n-p symmetric and for $b=0.25$ n-p asymmetric matter, 
we find no further meaningful  difference. 
It is worth to emphasize that the lower total density 
makes possible this non monotonic behavior.
This is produced the different behaviors with temperature
of  the functions 
$N_{\alpha} (T)$ and $M_{\alpha}(T)$ for given $V_0^{(0,1)}$ and $V_1^{(0,1)}$.
This becomes clearer with Figure 3, where $\rho = 0.5 \rho_0$.
As noticed above, the n-p asymmetry might amplify this non-monotonic behavior
although the difference is very small in the cases we show.

The same Skyrme parameterizations (and symbols) 
 are used in Fig. 3, where  $\cA_{0,1} (T)$ is exhibited 
for the still lower density $\rho = 0.5 \rho_0$ with: $b = 0.5$ 
(full solid symbols and $\times$) and $b=0.25$ (empty symbols and $+$).
Differently from all the results shown above the only Skyrme force that exhibit the non-monotonic behavior is SLyb for both $b=0.5$ and $b=0.25$.
Furthermore we notice that the stronger variation for all the forces 
(even if they are very small of the order of $.5$MeV) 
occur below $T \simeq 5$ MeV  or $T \simeq 10$MeV. 
Above these temperatures, the isovector polarizability variation is smaller.
The reason why the SGII and SLyb forces have the non-monotonic behavior at different densities is explained by  the different values of the functions 
$V_0^{(0,1)}$, in terms of the $t_0, t_3$ Skyrme parameters \cite{npa,prc}, 
and also $V_1^{(0,1)}$ (non zero).

For the range of lower nuclear densities some further conclusions can be extracted by comparing the figures 2 and 3, in particular for $b=.25$,
(which means $\rho_n = 1.25 \rho_p$). 
We notice that
the behavior of $\cA_{0,1} (T)$ is different depending on the effective force.
While the parameterizations SkSC 4 and SkSC 6 do not provide any different behavior (apart from an eventual overall total variation of $\cA_{0,1}(T)$), the forces SGII and SLyb present different trends for $\rho = .5 \rho_0$ (Fig. 3: empty squares, SGII, and empty diamonds, SLyb, respectively) and 
$\rho = .75\rho_0$ (Fig.2: full squares and full diamonds respectively).
The non-monotonic behavior appears for the SGII parameterization at 
$\rho=.75\rho_0$ while for the SLyb one it appears when $\rho= .5 \rho_0$.
As discussed in the beginning of the last section, because of the complicated form of the expression (\ref{20}) and of the behavior of the functions  $N(T)$ and $M(T)$,
the results from the polarizabilities with Skyrme forces are not always 
monotonic.
A suitable quantity for comparing the results from different forces, and even different methods,  is the total variation 
$\Delta A_{0,1} = A_{0,1} (T=20) -  A_{0,1} (T=0)$.
This quantity is (quite) small in all the works of the field, with small differences also due to the particular nuclear density under consideration 
being also seen also in other works using different approaches
\cite{moustakidis07,dean02-etal,xu06-etal,samaddar-etal} as analyzed
 in \cite{flb-aip}. 
One of the main outcomes of these comparisons goes along with the above remarks: the behavior of $\cA_{0,1}(T)$ might be different at $\rho_0$,
$0.75 \rho_0$ and $0.5 \rho_0$, eventually for still lower densities analyzed in other works.

In Figure 4, $\cA_{0,1} (T)$ is exhibited for a  density higher than 
$\rho_0$, i.e.
 $\rho = 1.33 \rho_0$ with $b=0$ 
(full symbols and $\times$) and $b=0.25$ (empty symbols and $+$).
The same kind of behavior found for 
$\rho = \rho_0$, in Figure 1, is present in Figure 4. 
The isovector polarizability very smoothly increases with temperature, 
although the variation is considerably smaller at high temperatures.

 This analysis suggests that the T-dependence of the s.e.c. is strongly dependent on the nuclear matter density $\rho$, and it is also suitable for 
shedding light on the nuclear effective interactions expected to be reliable with good predictive power.
Nevertheless we emphasize  that experimental data with temperature
 are very difficult to be extract unambiguously \cite{kowalski-etal}.

\section{ Summary}

To summarize we conclude that the bulk 
isovector symmetry energy does not vary considerably 
in a quite wide range of temperatures within the 
isovector polarizability
with Skyrme forces. 
As noticed in other works
the dependence of $\cA_{0,1}$ on the n-p density asymmetry is probably too strong.
This issue, on the other hand, does not modify  
the variation with the temperature meaningfully.
In the framework of the Skyrme-Hartree-Fock parameterization, 
 temperature effects arise from the functions $ M, N$ 
 given by expressions 
(\ref{densities-int}). They depend on the general properties of 
Skyrme-Hartree-Fock approach.
The variation of $\cA_{0,1} (T)$ depends strongly on the particular Skyrme force,
being always very smooth and small. 
The different contributions of the potential and kinetic parts of the symmetry
energy for fixed densities, by means of the functions $N(T)$ and 
$M(T)$, as well as the relative values of $V_0^{(0,1)}$ and $V_1^{(0,1)}$,
are responsible for these different results of each Skyrme interaction.
The variation (decrease) of  $N_{\alpha}(T)$ (which reduces to the 
n,p densities of states at zero temperature) 
with temperature is however the most relevant 
contribution for the results.
The larger variation of $A_{0,1} (T)$ occurs for 
$ \rho < \rho_0$, 
depending strongly on the effective force parameterization 
(for which $V_1^{(0,1)} \neq 0$, i.e. SGII and SLyb).
The final behavior is not always monotonic with temperature.
The differences in the overall variation of $\cA_{0,1}(T)$  for different densities below $\rho_0$, is seen also  
in other works using different approaches
\cite{moustakidis07,dean02-etal,xu06-etal,samaddar-etal}
as pointed out and compared in Ref.~\cite{flb-aip}.
The trends exhibited by the  
Skyrme parameterizations suggest that the eventual 
experimental knowledge of the behavior of the symmetry energy 
 with the temperature
will also contribute to better fine-tuning
of the effective interaction as well as to improving
its predictive power. This is clearer in the comparison between figures 2 and 3
(mainly for Skyrme forces SGII and SLyb).

\begin{acknowledgments}
This work was partially supported by IBEM and Ministry of Science and
 Technology of Brazil,
and FAPESP in the earlier
 stage of the investigation. 
The author thanks Sergio R. Souza 
for short discussions and a reading of the manuscript.

\end{acknowledgments}

\newpage

\begin{figure}[!b]
\epsfig{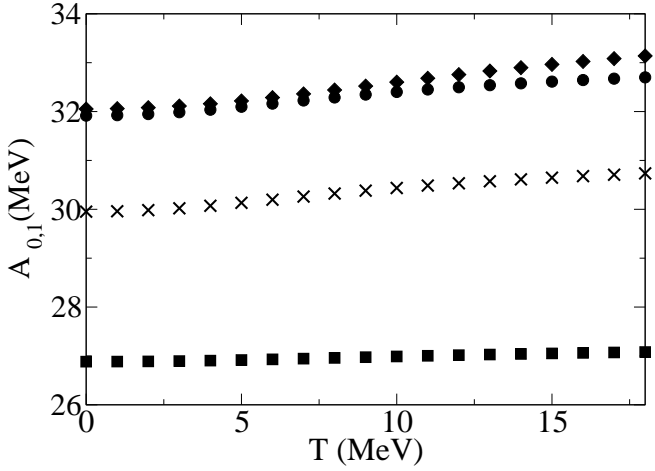}
       \caption{ 
The inverse of isovector polarizability,
 ${\cal A}_{0,1}$,
as a function of the temperature for  different
 Skyrme forces: SkSC 4 \cite{sksc} (circles), 
SGII \cite{sgii} (squares), SLyb
 \cite{sly} (diamonds), SkSC 6 \cite{sksc} ($\times$ or $+$).
For: ($\rho/\rho_0 = 1$ and  $b = 0$).
}
  \label{fig:1a}
\end{figure}

\begin{figure}[!b]
\epsfig{figure=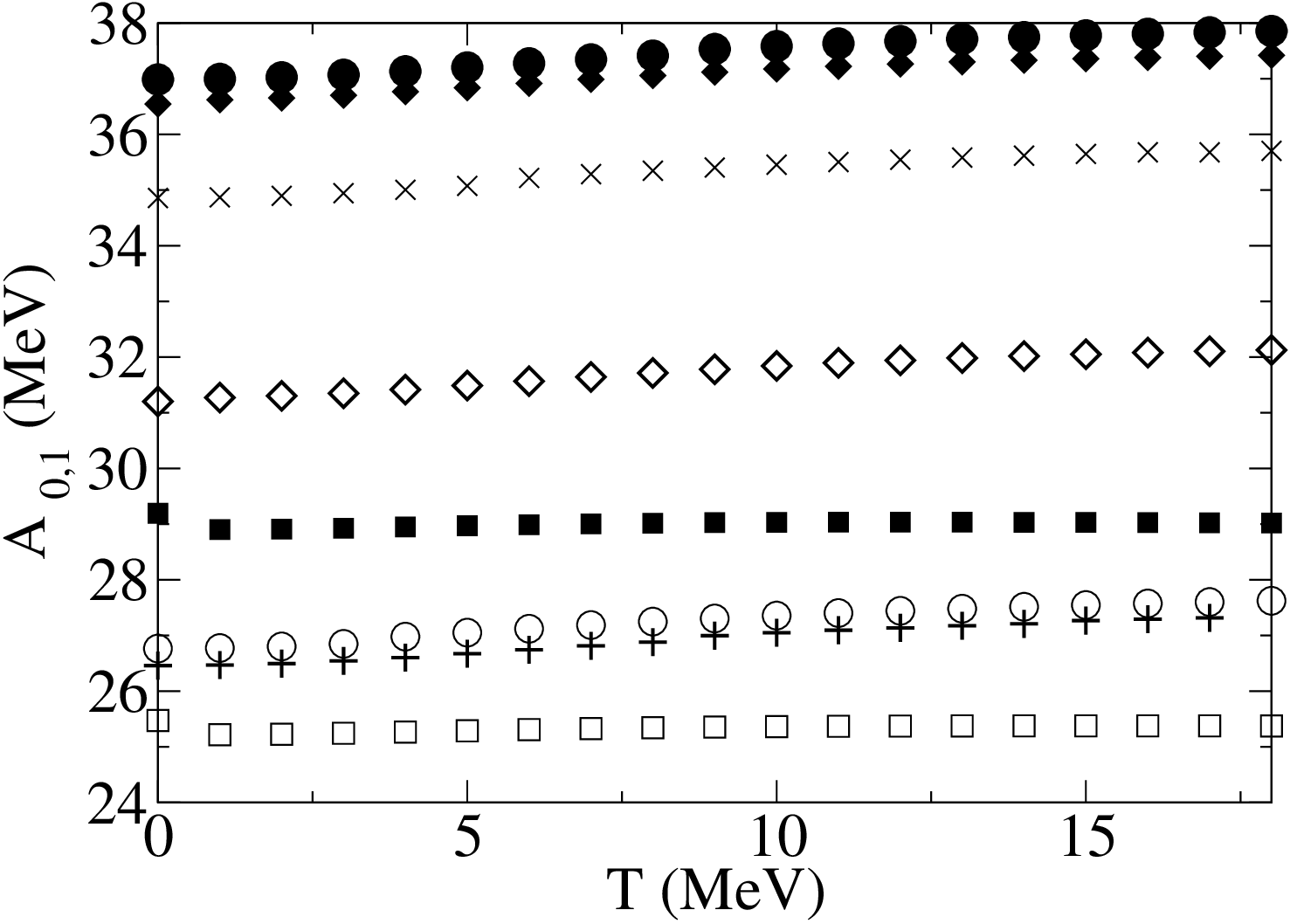,width=8.6cm}
       \caption{ 
The inverse of isovector polarizability,
 ${\cal A}_{0,1}$,
as a function of the temperature for the same
 Skyrme forces as in figure 1.
Considering: full symbols and symbol ($\times$) for 
($\rho/\rho_0 = 0.75$ and  $b = 0.25$)
and empty
symbols and $+$ for ($\rho/\rho_0 = 0.75$ and  $b = 0.$).
}
  \label{fig:2}
\end{figure}

\begin{figure}[!b]
\epsfig{figure=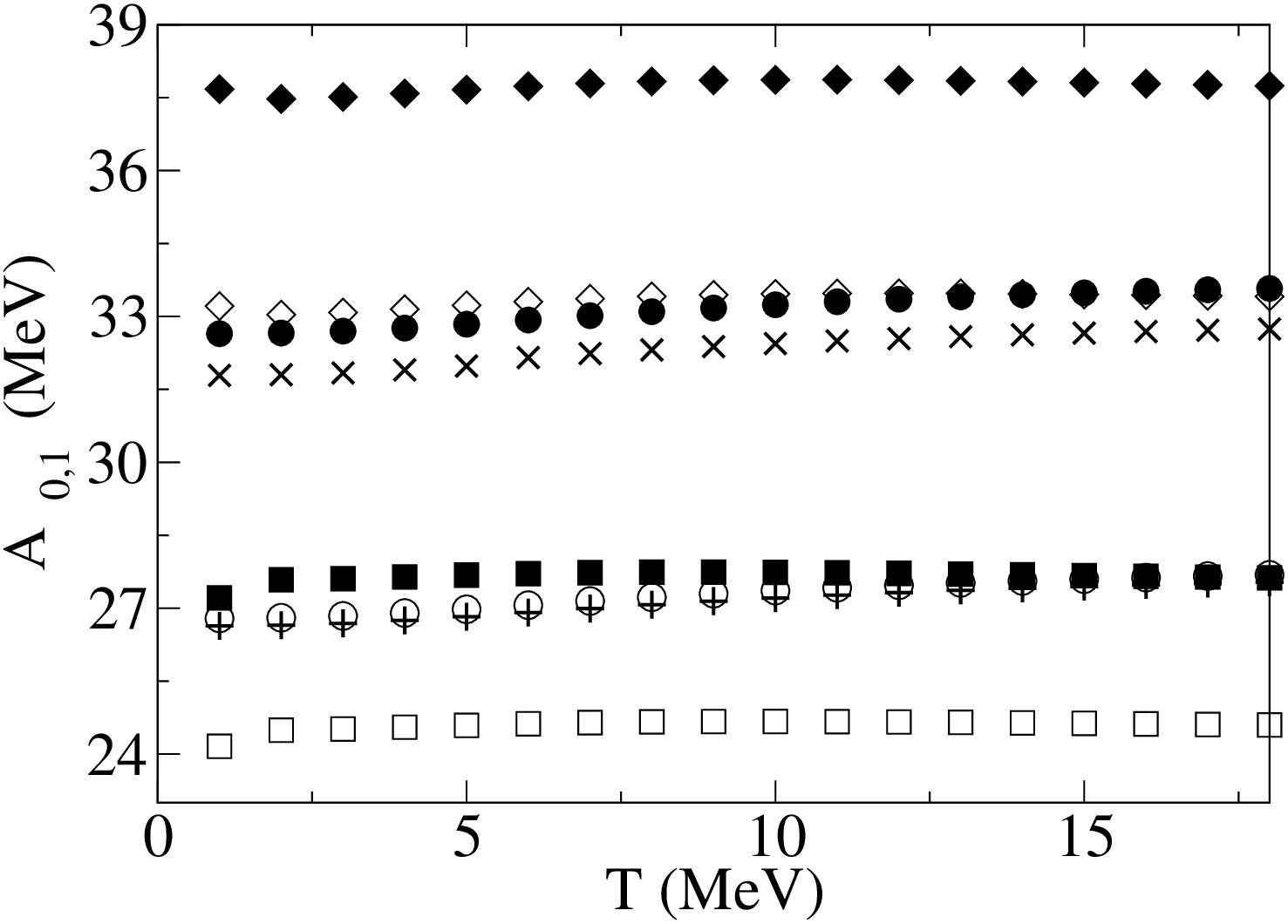,width=8.6cm}
  \caption{
The inverse of isovector polarizability,
 ${\cal A}_{0,1}$,
 as a function of the temperature for  the 
same Skyrme forces as figure 1.
Considering: full symbols and symbol $\times$ for 
($\rho/\rho_0 = 0.5$ and  $b = 0.5$)
and empty
symbols and $+$ for ($\rho/\rho_0 = 0.5$ and  $b = 0.25$).
}
  \label{fig:3}
\end{figure}

\begin{figure}[!b]
\epsfig{figure=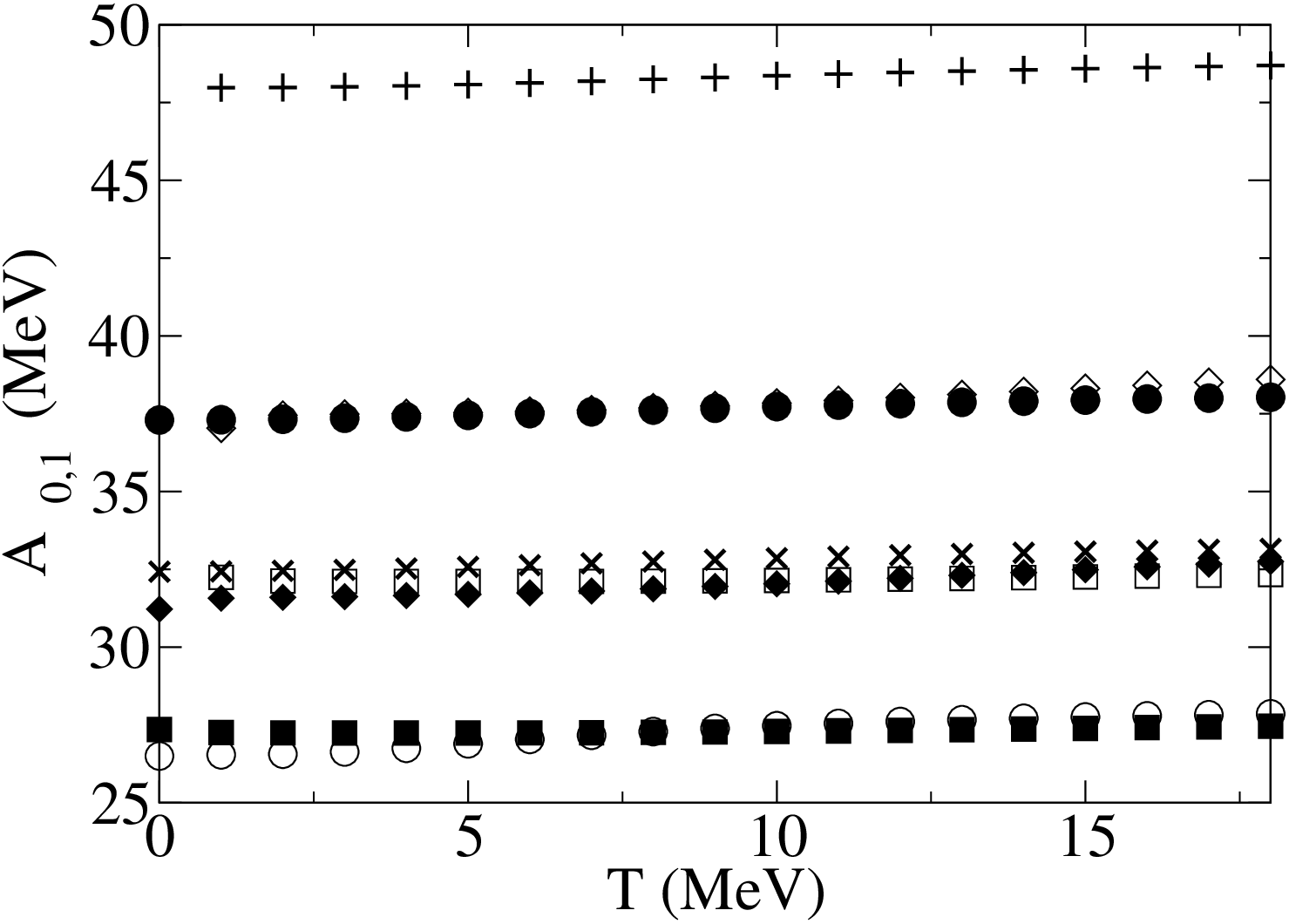,width=8.6cm}
  \caption{ 
The inverse of isovector polarizability,
 ${\cal A}_{0,1}$,
as a function of the temperature for  the 
same Skyrme forces as figure 1.
Considering: full symbols and symbol $\times$ for 
($\rho/\rho_0 = 1.33$ and  $b = 0.$)
and empty
symbols and $+$ for ($\rho/\rho_0 = 1.33$ and  $b = 0.25$).
}
  \label{fig:4}
\end{figure}


\begin{thebibliography}{11}




\bibitem{npa} F.L. Braghin,
 Nucl. Phys. {\bf A 665}, 13 (2000).



\bibitem{prc} F.L. Braghin,
 Phys. Rev.  {\bf C 71} 064303 (2005);  Erratum to be published in
Phys. Rev. C. (2009); F.L.B.
 Int. Journ. of Mod. Phys. {\bf E 12}, 755 (2003).


\bibitem{shetty-0704} D.V. Shetty, S.J. Yennello, G.A. Souliotis,
Phys.Rev. C 76, 024606 (2007); Erratum-ibid.C76, 039902 (2007). ArXiv:nucl-ex/0704.0471.


\bibitem{MSU-GSI} 
For example: W. Henning, Nucl. Phys. {\bf A 734} 654 (2004)
; J.A. Nolen, Nuc. Phys. {\bf A 734}, 661 (2004).
W.G. Lynch, Nuc. Phys. {\bf A 734}, 573 (2004). 
 "Isospin Physics in Heavy-Ion Collisions at Intermediate
Energies", Eds. Bao-An Li and W. Udo Schroeder, NOVA Science 
Publishers, Inc. (New York), (2001). 


\bibitem{BKB} B.-A. Li, C. M. Ko, W. Bauer, Int. Journ.
of Mod. Phys. {\bf E 7}, 147 (1998). 
 Bao-An Li, Nuc. Phys. {\bf A 734} 593 (2004).
B.-A. Li, C.B. Das, S. DasGupta, C. Gale,
Phys. Rev. {\bf C 69} 011603(R)  (2004).

\bibitem{NN2006} Several works presented in IX Conference on Nucleus Nucleus Collisions, Rio de Janeiro, Brazil, August 28 September 1, 2006. Ed. C.A. Bertulani, P.R.S. Gomes, M.S. Hussein, A.Szanto de Toledo, Nucl. Phys. {\bf A 787},(2007).


\bibitem{GAITANOS} T. Gaitanos, M. Di Toro, G. Ferini, M. Colonna,
H.H. Wolter, in {\it Proceedings of the XLII International Winter Meeting 
On Nuclear Physics}, Bormio (Italy) January-February, 2004;
ArXiv:nucl-th/0402041.
V. Grecco, M. Colonna, M. DiToro, F. Matera, Phys.Rev. C 67, 015203 (2003);
ArXiv:nucl-th/0205046. 
B. Liu, V. Greco, V. Baran, M. Colonna, M. DiToro, 
Phys. Rev. {\bf C 65} 045201, (2002).
F.L. Braghin, 
 Int. Journ. of Mod. Phys.   {\bf D 13}-7, 1267 (2004).

\bibitem{eos} P. Danielewicz, R. Lacey, W.G. Lynch,  Science {\bf 298},1592 (2002).
D.V. Shetty, S.J. Yennello, G.A. Souliotis, in
Proceedings of CAARI 2006, Forth Worth, Texas, Aug. 2006.
ArXiv:nucl-ex/0610019.


\bibitem{SCALING-FRAG} M.B. Tsang {\it et al} Phys. Rev. Lett. 86, 5023 (2001).
D.V. Shetty, {\it et al}, Phys.Rev. C 70, 011601(R)  (2004).
ArXiv:nucl-ex/0406008.
D.V. Shetty {\it et al}, nucl-ex/0401012.
 A.S. Botvina, O.V. Lozhkin, W. Trautmann, Phys. Rev. {\bf C 65},  044610 (2002).

\bibitem{ASTRO}  S. Reddy, M. Prakash, J.M. Lattimer, J.A. Pons,
  Phys. Rev. {\bf C 59}, (1999) 2888.
A.W. Steiner, M. Prakash, J.M. Lattimer, P.J. Ellis, 
Phys. Rept. 411, 325 (2005) 325. ArXiv:nucl-th/0410066.

\bibitem{kowalski-etal} S. Kowalski {\it et al}, Phys. Rev. C 75, 014601 
 (2007).
D.V. Shetty et al, arXiv:nucl-ex/0606032.
A. Le Fevre et al,for the ALADIN and INDRA Collaborations, Phys. Rev. Lett. 94, 162701 (2005); W.
Trautmann eat al, for the ALADIN and INDRA Collaborations, in
Proceedings of the IWM2005, Catania, Italy, Nov 2005. ArXiv: nucl-ex/0603027.


\bibitem{de-samaddar}
J. N. De and S. K. Samaddar,
 Phys.Rev. C 78, 065204 (2008).


\bibitem{NPA776} C.J. Horowitz, A. Schwenk,
Nucl. Phys. A 776, 55 (2006).




\bibitem{botvina-etal}  N. Buyukcizmeci, A.S. Botvina, I.N. Mishustin, R. Ogul, 
Phys.Rev. C 77, 034608 (2008);
arXiv:nucl-th/0711.3382.
A. Ono {\it et al}, Phys. Rev. C 70, 041604 (R) (2004). J. Iglio {\it et al}, 
Phys. Rev. C 74, 024605 (2006).

\bibitem{souza-etal}  S.R. Souza, M.B. Tsang, R. Donangelo, W.G. Lynch,
A.W. Steiner, Phys.Rev. C 78, 014605 (2008),
arXiv:nucl-th/0804.1352.



\bibitem{samaddar-etal} S.K. Samaddar, J.N. De, X. Vinas, M. Centelles, Phys. Rev. C 76, 041602(R) (2007).



\bibitem{moustakidis07} Ch.C. Moustakidis, Phys.Rev. C 76, 025805 (2007);
arXiv:nucl-th/0706.0698.


\bibitem{dean95-etal} D.J. Dean, S.E. Koonin, K. Langanke, P.B. Radha, Phys. Lett B 356, 429 (1995).


\bibitem{dean02-etal} D.J. Dean, K. Langanke, J.M. Sampaio, 
Phys.Rev. C 66, 045802 (2002); ArXiv:nucl-th/0203076.


\bibitem{xu06-etal} Jun Xu, Lie-Wen Chen, Bao-An Li, Hong-Ru Ma, 
Phys. Rev. C 75, 014607 (2007); ArXiv:nucl-th/0609035.


\bibitem{donati94-etal} P. Donati, P.M.Pizzochero, P.F. Bortignon, R.A. Broglia, Phys. Rev. Lett. 72, 2835 (1994)


\bibitem{MOLLER-NIX} 
For extremely refined mass formulae see for example:
P. Moeller, J.R. Nix, W.D. Myers and W.J. Swiatecki,
At. Data Nucl. Data Tables {\bf 59}, 185 (1995).


\bibitem{NEUTRONdensity} A. Trzci\'nska {\it et al}, Phys. Rev. 
Lett. {\bf 87}, 082501-1 (2001) and references therein.

\bibitem{BVA} F.L. Braghin, D. Vautherin and A. Abada,
Phys. Rev. {\bf C  52} 2504 (1995).

\bibitem{npa2001} F.L. Braghin, Nucl. Phys. {\bf A 696}, 413 (2001);
{\it Erratum} {\bf A 790}, 487(E) (2002).

\bibitem{flb-aip} F.L.Braghin, to appear in Proceedings of XXXI 
Brazilian Meeting on Nuclear Physics - 2008, AIP Conf. Proceedings, 
ed. by V. Guimar\~aes, J.R.B. de Oliveira.

\bibitem{rtfnb-02-07} F.L. Braghin, in {\it Proceedings of Brazilian Meeting on Nuclear Physics 2007}, ed. A. Suaide, Brazilian Physical Society, (2008).
F.L. Braghin, Braz. Journ. of Physics {\bf 33}, 255 (2003).


\bibitem{qmc-etc} P.A.M. Guichon, H.H. Matevosyan, N. Sandulescu, A.W. Thomas,
Nucl. Phys. {\bf A 772}, 1 (2006); arXiv:nucl-th/0603044.


\bibitem{cp-etal}
L. Brito, Ph. Chomaz,  D.P. Menezes, C. Providencia, Phys. Rev. {\bf C76},
044316 (2007);
arXiv:nucl-th/0704.3607


\bibitem{samaddar-etal-08} S.K. Samaddar, J.N. De, X. Vinas, M. Centelles, Phys. Rev. C 78, 034607 (2008).


\bibitem{sgii} N.Van Giai, H. Sagawa,
Phys Lett {\bf B 106}, 379, (1981).


\bibitem{sly} E. Chabanat {\it et al},  Nucl. Phys. {\bf A 627}, 
 (1997) 710.

\bibitem{sksc} M. Onsi, H. Przysiezniak and J.M. Pearson,
Phys. Rev. {\bf C 50}, (1994) 460.


\end{thebibliography}
\end{document}